\documentclass[prm, reprint, aps, amsmath, amssymb]{revtex4-2}
\usepackage{amsmath}

\usepackage{siunitx}
\usepackage[caption=false]{subfig}

\usepackage{lmodern} 
\usepackage[bookmarks, colorlinks=false]{hyperref}
\usepackage{mathtools}
\hypersetup{colorlinks=true, allcolors=blue}
\usepackage{tabularx}

\begin{document}
\title{Atomistic simulations of athermal irradiation creep and swelling of copper and tungsten in the high dose limit}

\author{Luca Reali}
\email{Luca.Reali@ukaea.uk, corresponding author}
\affiliation{United Kingdom Atomic Energy Authority, Culham Campus, Oxfordshire OX14 3DB, UK}

\author{Max Boleininger}
\email{Max.Boleininger@ukaea.uk}
\affiliation{United Kingdom Atomic Energy Authority, Culham Campus, Oxfordshire OX14 3DB, UK}

\author{Daniel R. Mason}
\email{Daniel.Mason@ukaea.uk}
\affiliation{United Kingdom Atomic Energy Authority, Culham Campus, Oxfordshire OX14 3DB, UK}

\author{Sergei L. Dudarev}
\email{Sergei.Dudarev@ukaea.uk}
\affiliation{United Kingdom Atomic Energy Authority, Culham Campus, Oxfordshire OX14 3DB, UK}

\begin{abstract}
Radiation creep and swelling are irreversible deformation phenomena occurring in materials irradiated even at low temperatures. On the microscopic scale, energetic particles initiate collision cascades, generating and eliminating defects that then interact and coalesce in the presence of internal and external stress. We investigate how copper and tungsten swell and deform under various applied stress states in the low- and high-energy irradiation limits. Simulations show that the two metals respond in a qualitatively similar manner, in a remarkable deviation from the fundamentally different low-temperature plastic behaviour of bcc and fcc. The deviatoric part of plastic strain is particularly sensitive to applied stress, leading to anisotropic dimensional changes. At the same time, the volume change, vacancy content and dislocation density are almost insensitive to the applied stress. Low- as opposed to high-energy irradiation gives rise to greater swelling, faster creep, and higher defect content for the same dose. Simulations show that even at low temperatures, where thermal creep is absent, irradiation results in a stress-dependent irreversible anisotropic deformation of considerable magnitude, with the orientation aligned with the orientation of applied stress. To model the high dose microstructures, we develop an algorithm that at the cost of about 25\% overestimation of the defect content is up to ten times faster than collision cascade simulations. The direct time integration of equations of motion of atoms in cascades is replaced by the minimisation of energy of molten spherical regions; multiple insertion of molten zones and the subsequent relaxation steps simulate the increasing radiation exposure.
\end{abstract}

\maketitle
\section{Introduction}
Quantifying and predicting radiation-induced deformations in nuclear reactors is a safety requirement, and it also enables alleviating the drawbacks of over-conservative design. Quantitative mesoscopic models for radiation swelling and dimensional changes provide input to large-scale simulations of reactor components and even full reactors.  So far, these models have been restricted to the treatment of isotropic swelling \cite{Dudarev2018, Reali2022, Ghazari2021, Reali2024}, even though the occurrence of anisotropic stress relaxation in structural materials under irradiation is well documented \cite{Matera2000,Luzginova2011}. Radiation creep can be beneficial, relaxing the local stress concentrations that otherwise initiate brittle failure. At the same time, in applications such as bolted joints, stress needs to be retained and its relaxation is undesirable \cite{Grossbeck1991}.  Mitigating the effects of low-temperature anisotropic swelling and stress relaxation is critically significant to the operation of actively cooled reactor components \cite{Feichtmayer2024}, where athermal defect accumulation and evolution acts as a source of macroscopic stress, through gradients of swelling \cite{Reali2022}. At the same time, the radiation-stimulated microstructural evolution provides a driving force for stress relaxation.

Radiation defects are created when an atom, impacted by an energetic neutron, ion, or electron, recoils with sufficient energy to initiate a cascade of atomic collisions that locally melt a volume up to about \SI{50}{\nano\meter}$^3$, sending microscopic heat and shock waves through the surrounding material \cite{Calder2010}. The subsequent recrystallisation leaves behind self-interstitial atoms and vacancies. Self-interstitials coalesce into dislocation structures, initially loops and then dislocation networks \cite{Gelles1981, Derlet2020, Wang2023, Wielunska2022, Boleininger2022}, whereas vacancies, if thermally mobile, coalesce into voids. Neutron irradiation also gives rise to transmutations, altering the chemical composition and leading to segregation and precipitation \cite{Durrschnabel2021, Was2007}.

Irradiation-induced deformation is tensorial in nature \cite{Bates1981} and is sensitive to the spectrum of recoil energies $E_{\rm{PKA}}$ of primary knock-on atoms (PKAs) \cite{Boleininger2023}, to the applied stress \cite{Bates1981} and to temperature \cite{Bhattacharya2020}. We introduce the dimensionless density of relaxation volumes of defects produced by irradiation \cite{Dudarev2018,Reali2022}
\begin{equation}
 \omega_{ij}=\omega_{ij}(T, \sigma_{ij}, E_{\textnormal{PKA}}, \textnormal{dose}, \textnormal{dose rate}),
\end{equation}
which acts as a source of radiation-induced deformation and involves contributions from both point defects and dislocations \cite{Boleininger2022}. 

Whereas the dose, expressed in the displacement per atom (dpa) units, and representing a measure of exposure of a material to radiation, is proportional to the kinetic energy transferred to the atoms in the material from the incident energetic particles \cite{Yang2021}, the \emph{effect} of the damage depends on how many defects survive the cascade recrystallisation, and how they interact and evolve. The dynamics of evolution of defects depends on their mobility, whether driven by stress or temperature, and the rate of generation of the defects themselves, i.e. the dose rate \cite{Boleininger2023}. 

The trace of the relaxation volume density, ${\omega_{kk}=\omega_\mathrm{xx}+\omega_\mathrm{yy}+\omega_\mathrm{zz}}$, is a measure of local irradiation-induced volume change, also known as \emph{swelling}. The same term is sometimes used when referring to void swelling that occurs in heavily irradiated materials in a certain range of temperatures, where void swelling can reach values as high as tens of percent \cite{Garner2020}. From the above definition, swelling is a scalar quantity. At the same time, materials exposed to irradiation not only expand volumetrically but also exhibit anisotropic dimensional changes. For instance, the exposure of stainless steel to neutron irradiation under applied stress produces anisotropic deformation and, surprisingly, as a function of dose this deformation remains anisotropic even after the applied stress is removed \cite{Garner2009}. 

The deviatoric part of $\omega_{ij}$ describes how strongly the defect and dislocation microstructure polarises in response to the applied stress. Microstructural polarisation commonly occurs due to the anisotropic stress-induced bias acting during nucleation \cite{DaFonseca2023} and the subsequent collective evolution of defects and dislocations  \cite{Feichtmayer2024,Matthews1988,Onimus2020, Was2007}. At low temperatures, swelling is low and does not exceed {1\,\%} \cite{Boleininger2022}. Its gradients can still induce substantial macroscopic stress of order $\mu \omega_{kk}$ in reactor components, where $\mu $ is the shear modulus \cite{Reali2022,Dudarev2018}. 

Anomalously rapid radiation creep, occurring at low temperature and fundamentally different from thermal creep, is well documented. Creep, occurring under uniaxial loading in tungsten (W) and molybdenum exposed to irradiation, was observed even at \SI{20}{\kelvin} \cite{Ponsoye1971, pouchou1975low}, and the experimentally measured radiation creep rates in stainless steel at \SI{60}{\celsius} are higher than at \SI{330}{\celsius} \cite{Grossbeck1991}. Stress relaxation in Ni and Inconel is more pronounced at \SI{330}{\kelvin} than at \SI{570}{\kelvin} \cite{Causey1980, Griffiths2019}. Motivated by these observations, we undertake a computational study of the effects of stress and recoil energy on microstructure and deformation in the high-dose, low-temperature limit, where the thermally activated migration of defects is either very slow or inactive \cite{Derlet2020}. 

\emph{In silico} simulations of radiation damage have an established track record: molecular dynamics (MD) simulations have been extensively applied to investigate how defects form in the very low dose limit \cite{Nordlund2019}. Collision cascades simulated in strained W exhibited enhanced defect production under tension, in the large strain limit \cite{Wang2016}. Simulations of high-dose microstructures have recently become feasible through the use of Frenkel pair (FP) accumulation models \cite{Chartier2019, Derlet2020}, overlapping cascade simulations \cite{Boleininger2023}, and a combination of the two methods \cite{Mason2021}. The latter two algorithms are computationally demanding while the former two methods overestimate the predicted defect concentrations \cite{Mason2020}.

The effect of applied stress in the high-dose regime was investigated using FP accumulation simulations in iron under high uniaxial stress approaching \SI{1}{\giga\pascal}. Interstitial-type defects were found to polarise the microstructure in response to the applies stress, whereas the total dislocation and vacancy content remains largely unaffected by stress \cite{Stefanescu2023}. In a recent study of strain- rather than stress-controlled radiation creep of tungsten at room temperature, overlapping cascade simulations accurately reproduced the experimentally observed stress relaxation \cite{Feichtmayer2024}, exhibiting good quantitative agreement between predictions and observations even at high radiation exposure.

A computational tool, aspiring to bridge the gap between the dislocation representation of microstructure and macroscopic stress analyses in a material not exposed to irradiation, is the crystal plasticity finite element (CPFE) method  \cite{Nguyen2021}, parameterised using atomistic simulations \cite{Bertin2023} and treating the glide and climb of dislocations. If $\mathbf{x}$ is a coordinate of a body deformed from a reference coordinate $\mathbf{x}_0$, the mapping between the two is established by the deformation gradient tensor $\mathbf{F}_{ij}=\partial\mathbf{x}_i/\partial{\mathbf{x}_0}_j$, which is an observable quantity \cite{Mason2024}. Using the chain differentiation rule, $\mathbf{F}$ can be decomposed into a product of an elastic and a plastic term \cite{Yu2021}
\begin{equation} \label{eq:CPdecomposition}
    \mathbf{F}=\mathbf{F}^\mathrm{e}\mathbf{F}^\mathrm{p}.
\end{equation}
In body-centred cubic (bcc) metals, the plastic contribution is constrained by the rate of thermally activated glide of screw dislocations. This results in that the shear deformation rate entering $\mathbf{F}^\mathrm{p}$ is characterised by the temperature dependence of the Arrhenius form $\Dot{\gamma}\propto\exp(-\Delta G/k_\textnormal{B}T)$, where $\Delta G$ is the kink-pair nucleation energy \cite{Yu2021, Monnet2013}. Therefore, $\mathbf{F}^\mathrm{p}\rightarrow {\bf I}$ if $T\rightarrow0$. In face-centred cubic (fcc) metals, the Arrhenius temperature factor does not enter the dislocation mobility equations \cite{Zhao2007}, resulting in a very different pattern of plastic behaviour at low temperatures.

The plastic behaviour of materials profoundly changes under, or following, irradiation \cite{Arsenlis2004}, depending on the material and irradiation conditions, because of the formation of obstacles to dislocation glide such as dislocation loops, Laves phase inclusions, or the stacking fault tetrahedra (SFT). Crystal plasticity (CP) models, taking into account irradiation effects, were applied to modelling straining of W \cite{Yu2021} and steel \cite{Yu2022}, as well as irradiation growth in Zr \cite{Patra2017}, where rate equations were applied to describe void and dislocation loop nucleation and growth. The CP formalism is not able to describe the low-temperature radiation creep of bcc metals and its observed saturation as a function of dose \cite{Ponsoye1971, Feichtmayer2024}. 

Below, we explore the irradiation-driven deformations in the limit where conventional plasticity concepts do not apply. At low temperature, deformation is driven purely by the energetic particle impact events, and the rate of deformation is related to the irradiation dose rate. The fundamental origin of deformation in this case is the stochastic accumulation and polarisation of eigenstrain of radiation defects and dislocations \cite{Reali2022,Feichtmayer2024}. 

\section{Methods: the molten spheres algorithm}
\begin{figure}[t]
  \includegraphics[width=0.95\columnwidth]{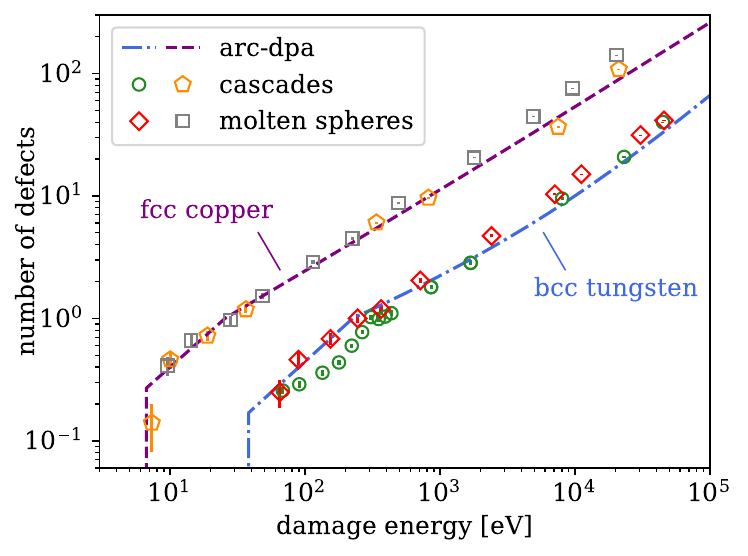} %
    \caption{Numbers of defects produced by individual collision cascades, parametrised using the arc-dpa model \cite{Nordlund2018,Yang2021}. Defect numbers predicted by melting and then minimizing the energy of molten spherical regions satisfactorily reproduce the results of significantly more computationally intensive direct cascade simulations across a broad range of PKA energies up to the cascade splitting threshold \cite{DeBacker2016}. Standard error bars are very narrow and visible only inside the markers except for the lower energies.}
    \label{fig:N_def} 
\end{figure}

When an energetic particle impact initiates an atomic recoil, referred to as the primary knock-on atom (PKA) \cite{Calder2010}, the material in a small region around the PKA \emph{melts} and subsequently recrystallises, leading to the formation of radiation defects. Hence, we expect to be able to capture most of the relevant defect formation processes by randomly selecting the points of origin of the PKAs, and then melting and athermally recrystallising the corresponding spatial regions through energy minimisation. We simulated radiation damage of tungsten and copper at low temperatures in LAMMPS \cite{Plimpton1995} using a bespoke molten spheres algorithm involving the following steps: 
\begin{enumerate}
    \item ``Melt'' $N$ randomly located spherical regions by replacing the inside atoms with the same number of atoms taken from a molten configuration.
    \item Perform energy minimisation to ``cool down'' the molten regions.
    \item Repeat until the target dose is reached.
\end{enumerate}
The number of surviving FPs in the wake of an \emph{isolated} cascade can be evaluated using the Yang-Olsson \cite{Yang2021} modified arc-dpa model \cite{Nordlund2018}, which is parametrised using full cascade simulations. To confirm the validity of our algorithm, in Fig.~\ref{fig:N_def} we compare the predicted number of FPs: (i) after inserting and relaxing one molten sphere, (ii) after a single conventional collision cascade using data from \cite{Boleininger2023}, (iii) using the modified arc-dpa model. For the former two cases, independent simulations were repeated several times for statistical significance. The agreement between the data is very good over a broad range of energies up to the cascade splitting threshold \cite{DeBacker2016}.

To vary the effective PKA energy corresponding to a molten sphere, the radius of the molten region needs to be appropriately chosen. To calculate it, we recall that the number of remaining FPs depends on the damage energy $T_{\rm{d}}$, which is the part of $E_{\rm{PKA}}$ that is not lost to electronic excitations \cite{Lindhard1963}. $T_{\rm{d}}/E_{\rm{PKA}}\sim0.9$ for \SI{100}{\electronvolt} and $\sim$0.8 for \SI{10}{\kilo\electronvolt} cascades. If $T_{\rm{d}}$ is transferred from a single atom of atomic volume $\Omega_0$ to molten atoms inside a sphere of radius $R$, each acquiring an effective melting energy $E_{\rm{melt}}$, the expression for determining $R$ reads
\begin{equation}\label{eq:Td_R}
    T_{\rm{d}}=\frac{4\pi}{3}\frac{R^3}{\Omega_0}E_{\rm{melt}}.
\end{equation}
$E_{\rm{melt}}$ is the only parameter of the model and can be determined from inexpensive single cascade simulations following the procedure outlined in Ref.~\cite{Boleininger2023}. For W and Cu, $E_{\rm{melt}}$ is 2.70 and \SI{0.63}{\electronvolt}, respectively \cite{Boleininger2023}. Absent a bespoke parametrisation, an estimate for this quantity is $E_{\rm{melt}}=c_{\rm{p}}T_{\rm{m}}+h_{\rm{m}}$, given the specific heat, melting temperature and the enthalpy of fusion of the material. The quality of agreement with full cascade simulations is very good. A difference is that we find self-interstitials at the core of the melt surrounded by vacancies at the periphery of the molten zone, whereas the opposite is found in cascade simulations \cite{Nordlund2018}. Although in this approximation we neglect the occurrence of shock waves, thermally-activated diffusion \cite{Calder2010}, and thermal spikes \cite{vineyard1976thermal}, we are still able to capture the key features of irradiation exposure, amounting to the creation and annihilation of crystal defects through the rapid quenching of the molten structures \cite{Nordlund1997}. 

To reach high irradiation dose, subsequent minimisation step were performed, each time inserting $N$ molten spheres at random positions ensuring that their surfaces were at least \SI{15}{\angstrom} apart. Each region contributed a damage energy $T_{\rm{d}}$, and the corresponding radii followed from Eq.~\eqref{eq:Td_R}. According to the commonly used NRT-dpa damage model \cite{Norgett1975}, this corresponds to an increment in dose $\phi$ at every algorithmic step of
\begin{equation}\label{eq:NRT}
    \Delta\phi=N\frac{0.8T_{\rm{d}}}{2E_{\rm{d}}N_{\rm{at}}},
\end{equation}
where $E_{\rm{d}}$ is the threshold displacement energy and $N_{\rm{at}}$ is the number of atoms. $E_{\rm{d}}$ greatly depends on the direction of the recoiling PKA \cite{Banisalman2017} and there is some uncertainty introduced by reducing the treatment to an isotropic spherical approximation. $N$ was adjusted so that $\Delta\phi$ was between 0.1 and {0.3\,mdpa} in all cases. $E_{\rm{d}}$ for Cu is lower than the experimental value \cite{Kenik1975}. It was selected based on our MD simulations \cite{Boleininger2023} to be appropriate for the interatomic potential that was used. Note that, in the case of low-energy W irradiation, $T_{\rm{d}}$ equals $E_{\rm{d}}$. According to the NRT-dpa model \cite{Norgett1975, Yang2021}, this is exactly the energy where the damage rate exhibits a discontinuity from 0 to 1 Frenkel pairs per recoil. Since this is an oversimplification of the model, see for example the MD results of \cite{Yang2021}, we apply Eq.~\eqref{eq:NRT} in this case as well, effectively obtaining the intermediate value of 0.4 Frenkel pairs per recoil. Therefore, due to both the anisotropy of $E_{\rm{d}}$ and the approximate nature of the NRT-dpa measure \cite{Nordlund2018}, some uncertainty is involved in the quantification of dose in the simulations. This is not a feature of our method, but of cascade simulations as well. It does, however, only affect the dose at which simulations yield their results and not the results themselves. 

The algorithm is a physically realistic extension of the FP accumulation method \cite{Chartier2019, Derlet2020}, with the additional feature of being able to account for different PKA energies, which are known to have a significant effect on microstructural evolution in high-dose simulations \cite{Boleininger2022}. This method simplifies the cascade-driven damage production process in at least two important ways. The collision cascade step is extremely simplified, and defect reorganisation due to the localised cascade heating is suppressed. This includes both thermal effects as well as shock waves propagating away from the cascade even if melting does not occur \cite{Heredia2021}. As a result, some secondary effects such as cascade-induced unpinning of dislocations may not be treated \cite{Khiara2020}. Any thermal evolution occurring between the cascade events is also neglected. The former can be removed at the expense of running explicit molecular dynamics, implying some significant computational cost. The latter can be addressed through the application of accelerated molecular dynamics methods, however, bridging MD and technologically relevant time-scales still remains exceptionally difficult. The molten spheres method approaches athermal irradiation conditions, where the dose rate is high compared to the diffusion time scales \cite{Boleininger2023}, since no time propagation is allowed except for the effective time of the minimisation algorithm. It is reasonable to compare the results to other athermal computational simulations or experimental conditions where there is negligible long-term defect recombination. This can be due to low temperature, impurities pinning the dislocations, or a combination of both. 

A significant advantage offered by the algorithm is that it is faster than cascade simulations by about a factor of 10, enabling us to reach the dose of {1\,dpa} for about 30 simulations for this study, while using 16-million atom simulation cells required to described complex defect and dislocation microstructures. If Fig.~\ref{fig:N_def} was concerned with primary radiation damage, we also performed a series of tests to prove that the microstructures generated by the simulations are qualitatively and quantitatively representative of high-dose irradiation at low temperature. In comparison with full cascade simulations, the algorithm overestimates the high-dose defect content by about 25\%, as detailed in the Supplementary Material. Finally, we note that even if we focused here on single-energy irradiation, the method is readily generalisable to energy-resolved neutron spectra by drawing the radii from a distribution corresponding to that of the actual spectrum of recoils through Eq.~\eqref{eq:Td_R}. An example from ion-irradiation is given in the Supplementary Material.

Our objective is the assessment of the response of tungsten and copper to a simultaneous exposure to radiation and constant external applied stress. The cubic unit cell was aligned with a periodic supercell of side length of about \SI{60}{\nano\meter} consisting of $200^3$ and $159^3$ unit cells for W and Cu, containing about 16 million atoms. The triclinic supercells were allowed to freely change volume \emph{and} shape, under periodic boundary conditions. The interatomic potentials were taken from Refs.~\cite{Mason2017, Ackland1990}. We considered low- and high-energy irradiation, namely \SI{90}{\electronvolt} and \SI{7}{\kilo\electronvolt} effective PKA energy events in W, and \SI{100}{\electronvolt} and \SI{5}{\kilo\electronvolt} events in Cu. The six stress state types explored in this study are (i) the state of zero stress, (ii) uniaxial tension, (iii) uniaxial compression, (iv) hydrostatic tension, (v) pure shear with one non-zero stress component, and (vi) pure shear with identical three off-diagonal stress components. The magnitude of stress tensor components in all the above cases is chosen to be 500~MPa for W and 100~MPa for Cu, corresponding to approximately 0.1\% of their respective Young's moduli. 

To analyse the irradiation-induced deformations, we evaluate the supercell strain tensor $\varepsilon_{ij}$ using the definitions valid beyond the infinitesimal strain approximation, detailed in the Supplementary Material. $\varepsilon_{ij}$ is the total strain, which is the sum of volume-average relaxation volume density of irradiation defects $\omega_{ij}$ plus the volume-average elastic strain related to the applied stress by Hooke's law $e_{ij}=C_{ijkl}\sigma_{kl}$. For a constant applied stress, $e_{ij}$ remains constant throughout a simulation, while $\omega_{ij}$ evolves as a function of radiation exposure. 

We quantify swelling of the simulation cell by the trace of the total strain tensor $\varepsilon_{kk}$ averaged over the cell, and the volume-conserving creep deformation by the von Mises strain, also computed from the cell-average total strain as $\varepsilon_{\textnormal{vM}}=\sqrt{\frac{3}{2}\left[\varepsilon_{ij}\varepsilon_{ji}-\frac{1}{3}(\varepsilon_{kk})^2 \right]}$. In what follows, $\varepsilon_{ij}-\frac{1}{3}\varepsilon_{kk}\delta_{ij}$ is referred to as the deviatoric part of $\varepsilon_{ij}$, and $\frac{1}{3}\varepsilon_{kk}\delta_{ij}$ as the isotropic hydrostatic part of strain. Similar definitions were used in atomistic \cite{Stefanescu2023} and CP \cite{Yu2022} simulations. Below we explore the evolution of $\varepsilon_{ij}$ seeing how, as the dose increases, $\omega_{ij}$ rapidly becomes its dominant part. Towards the end, we also extract the volume-average values of $\omega_{ij}$ by relaxing the simulation cells to zero average stress or, equivalently, to zero average elastic strain.

\section{Results and Discussion}
\subsection{Creep and swelling under stress and irradiation}
Over the course of exposure to radiation, the volume and the shape of the simulation cell change. For instance, the elastic strain tensor $e_{ij}$ of copper under uniaxial tension prior to radiation exposure is 
\begin{equation}\label{eq:el_tension}
\begin{bmatrix}
-0.063 & 0 & 0\\
0 & -0.063 & 0\\
0 & 0 & 0.153
\end{bmatrix}\%.    
\end{equation}
After 1\;dpa of low-energy irradiation, the volume-average total strain in the cell is now
\begin{equation}\label{eq:1dpa_tension}
\begin{bmatrix}
0.54 & 0.07 & 0.04\\
0.07 & 0.58 & -0.01\\
0.04 & -0.01 & 1.11
\end{bmatrix}\%.
\end{equation}
This total strain, which is a sum of elastic and eigenstrain tensors, can be represented by a sum of isotropic and deviatoric parts,
$$
\begin{bmatrix}
0.74 & 0 & 0\\
0 & 0.74 & 0\\
0 & 0 & 0.74
\end{bmatrix}\%+
\begin{bmatrix}
-0.20 & 0.07 & 0.04\\
0.07 & -0.24 & -0.01\\
0.04 & -0.01 & 0.37
\end{bmatrix}\%.
$$
Similarly, under pure shear loading, a simulation cell describing copper exposed to irradiation, evolves from its un-irradiated state of purely elastic strain 
$$
\begin{bmatrix}
-0.001 & 0 & -0.066\\
0 & 0.001 & 0\\
-0.066 & 0 & -0.001
\end{bmatrix}\%
$$
to a state at 1\;dpa, containing now both elastic and eigenstrain components, here represented by isotropic and deviatoric strains 
$$
\begin{bmatrix}
0.73 & 0 & 0\\
0 & 0.73 & 0\\
0 & 0 & 0.73
\end{bmatrix}\%+
\begin{bmatrix}
0.16 & 0.06 & -1.17\\
0.06 & -0.52 & -0.09\\
-1.17 & -0.09 & 0.35
\end{bmatrix}\%.
$$
These examples show that irradiation gives rise to creep strains many times the elastic ones; while the applied stress has little effect on the hydrostatic isotropic strain, it generates a \emph{deviatoric} component that evolves as a function of irradiation exposure, remaining anisotropically aligned with the elastic strain. This was systematically observed for both W and Cu across all tested irradiation conditions. 

\begin{figure*}[t]
\subfloat[]{%
  \includegraphics[width=0.62\textwidth]{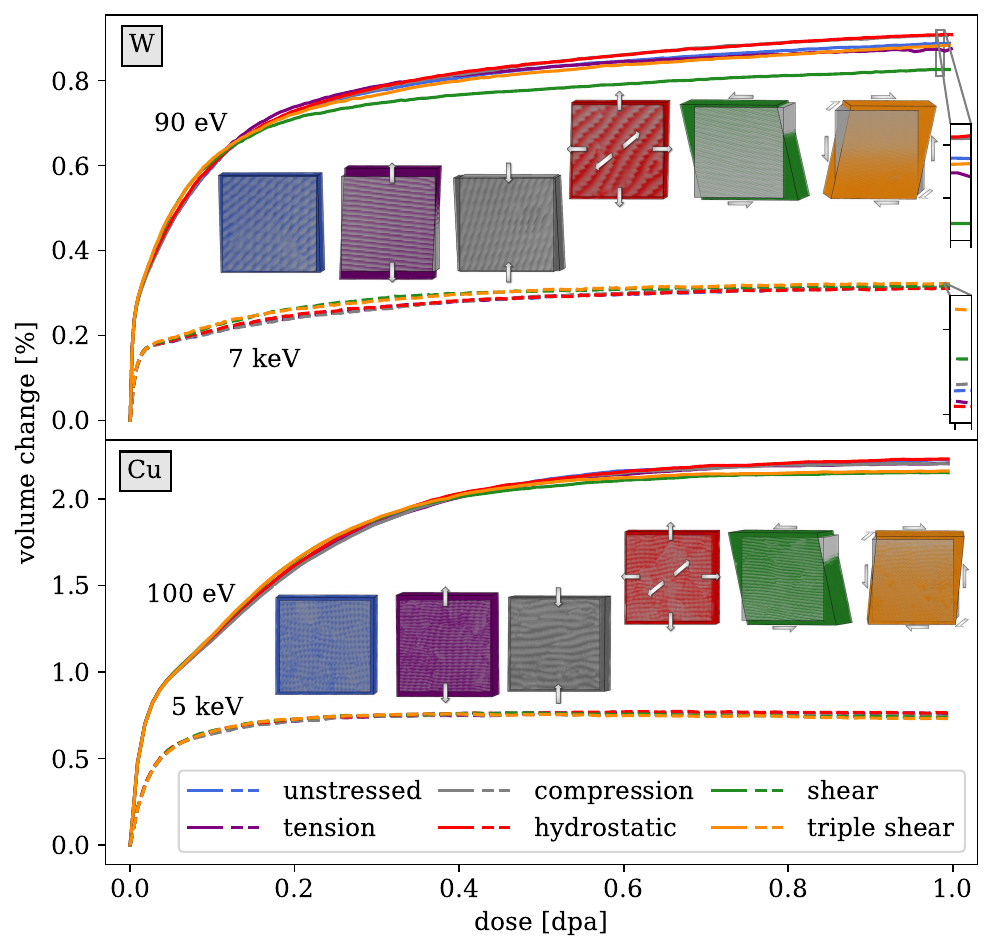}\label{fig:delta_V_a} %
}
\subfloat[]{%
  \includegraphics[width=.355\textwidth]{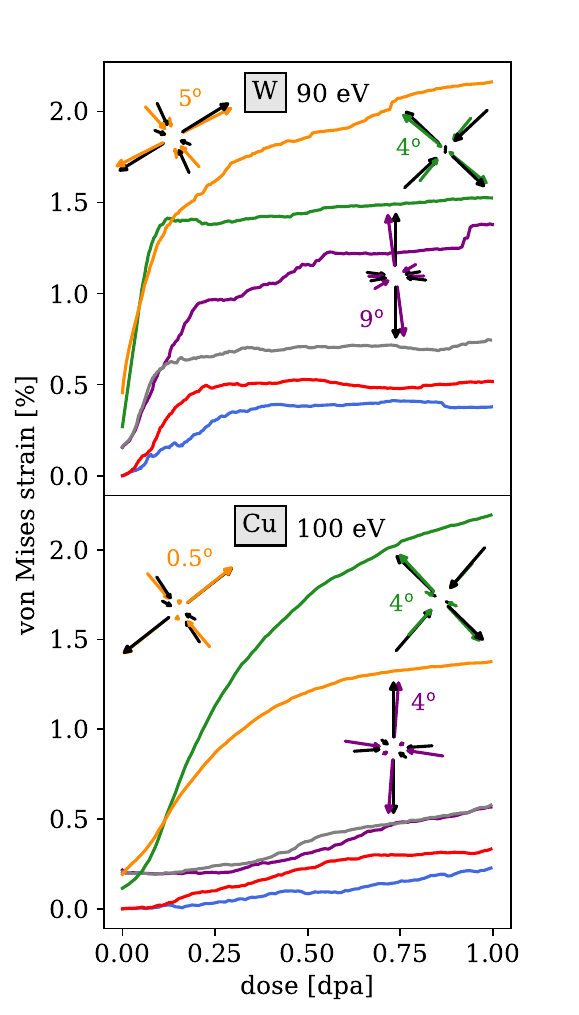}\label{fig:delta_V_b} %
}\hfill

    \caption{(a) Volume increase $\Delta V/V_0$ under irradiation and applied stress. Dimensional changes are illustrated by superimposing the actual atomistic configuration at {1\,dpa}, for 90 and 100\;eV irradiation in W and Cu, onto unirradiated supercells, with supercell deformations magnified by a factor of ten for demonstration. The figure shows that the hydrostatic isotropic strain $\varepsilon_{\textnormal{hyd}}\sim\Delta V/(3V_0)\delta_{ij}$ is almost unaffected by the applied stress. The patterns of behaviour exhibited by W and Cu are qualitatively similar, with high-energy recoils producing lower swelling. The von Mises strain curves in (b) show a strong effect of applied stress on irradiation creep with respect to the initial elastic offset. To illustrate the polarisation of swelling, we indicate the normalised eigenvectors of the deviatoric strain tensor and indicate the degree of misalignment between the principal straining direction at 0 (black) and {1\,dpa} (in colour). Even after very large increases in $\varepsilon_{\textnormal{vM}}$ during irradiation, the anisotropy of strain remains closely aligned with the applied stress. (a) and (b) show the hydrostatic and deviatoric components of the \emph{same} total strain tensor $\varepsilon_{ij}$.}
    \label{fig:delta_V} 
\end{figure*}

A typical pattern of changes of cell volume and shape occurring under irradiation is illustrated in Figs.~\ref{fig:delta_V}a-b. The response of the two materials to irradiation and stress is qualitatively similar, although Cu exhibits about twice as much swelling. Swelling saturates after exposure to 0.3-0.5 dpa, whereas creep still continues in Cu at {1\,dpa}. Volumetric swelling is higher for lower recoil energies \cite{Boleininger2023}, and it is nearly independent of stress. On the other hand, the deviatoric part of the total strain tensor depends sensitively on the applied stress. The irradiation-induced total von Mises strain increases by up to an order of magnitude over the dose interval of {1\;dpa}, with the applied shear stress having a particularly strong effect. 

The effect of applied stress on the irradiation creep strain goes beyond the magnitude of the von Mises invariant; by comparing the misorientation between the eigenvectors of the initial elastic strain tensor with those of the {1\;dpa} strain, e.g. the strain states shown in Eqs.~\eqref{eq:el_tension} and \eqref{eq:1dpa_tension}, we see that the irradiation-induced shape change remains anisotropically aligned with the initial elastic deformation. This demonstrates that externally-applied stress induces a polarisation in the tensorial relaxation volumes of defects produced by irradiation. Although the maximum principal strain along the transformation pathway, linking tensor \eqref{eq:el_tension} to \eqref{eq:1dpa_tension}, increases seven-fold, the corresponding eigenvectors barely change: the material rotates by just over 4\textdegree\ from the initial uniaxial loading direction $(0, 0, 1)$ to direction $(0.074, -0.016,  0.997)$. The corresponding eigenvectors are shown in black and purple in Fig.~\ref{fig:delta_V}b, and similar a pattern of evolution is observed in all the eight examples where the initial elastic eigenvectors are shown using the suitably colour-coded {1\;dpa} eigenvectors for tension, shear and compression. Overall, the lower energy recoils corresponding to lower $E_\textnormal{PKA}$ produce higher strains. We emphasize that this effect is particularly strongly pronounced in the athermal limit: already at \SI{300}{\kelvin} more defects recombine at the moment of production, and the surviving defect fraction can be as low as 0.05-0.30, depending on the material, temperature, and $E_\textnormal{PKA}$ \cite{Zinkle1993, Zinkle2023}. Given the absence of thermal phenomena contributing to defect recombination, our findings represent the upper limit of the effect of defect production for a given damage energy.

\begin{figure*}[p]
\subfloat[]{%
  \includegraphics[width=.95\textwidth]{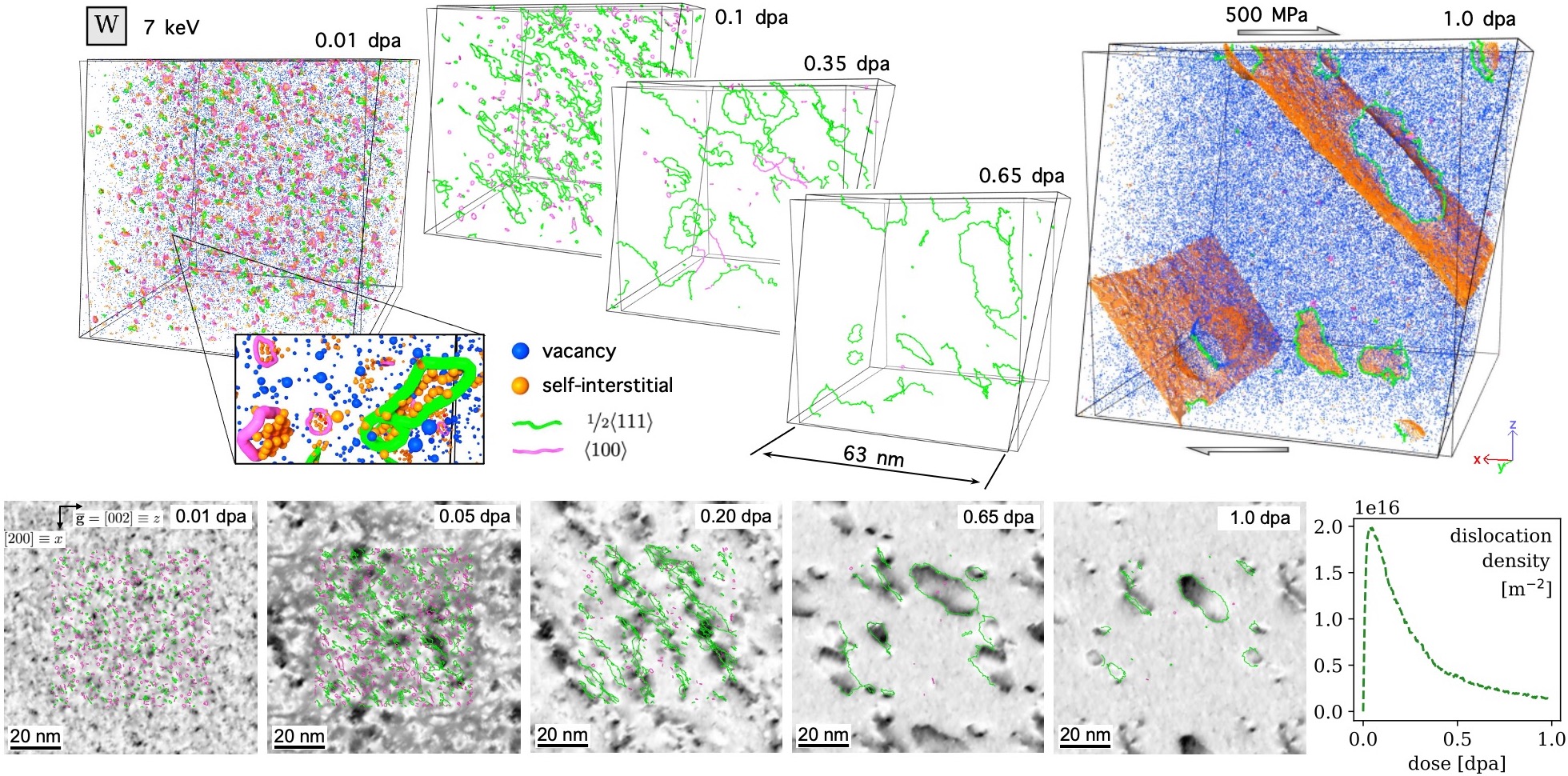}\label{fig:disloc_a} %
}\hfill
\subfloat[]{%
  \includegraphics[width=.95\textwidth]{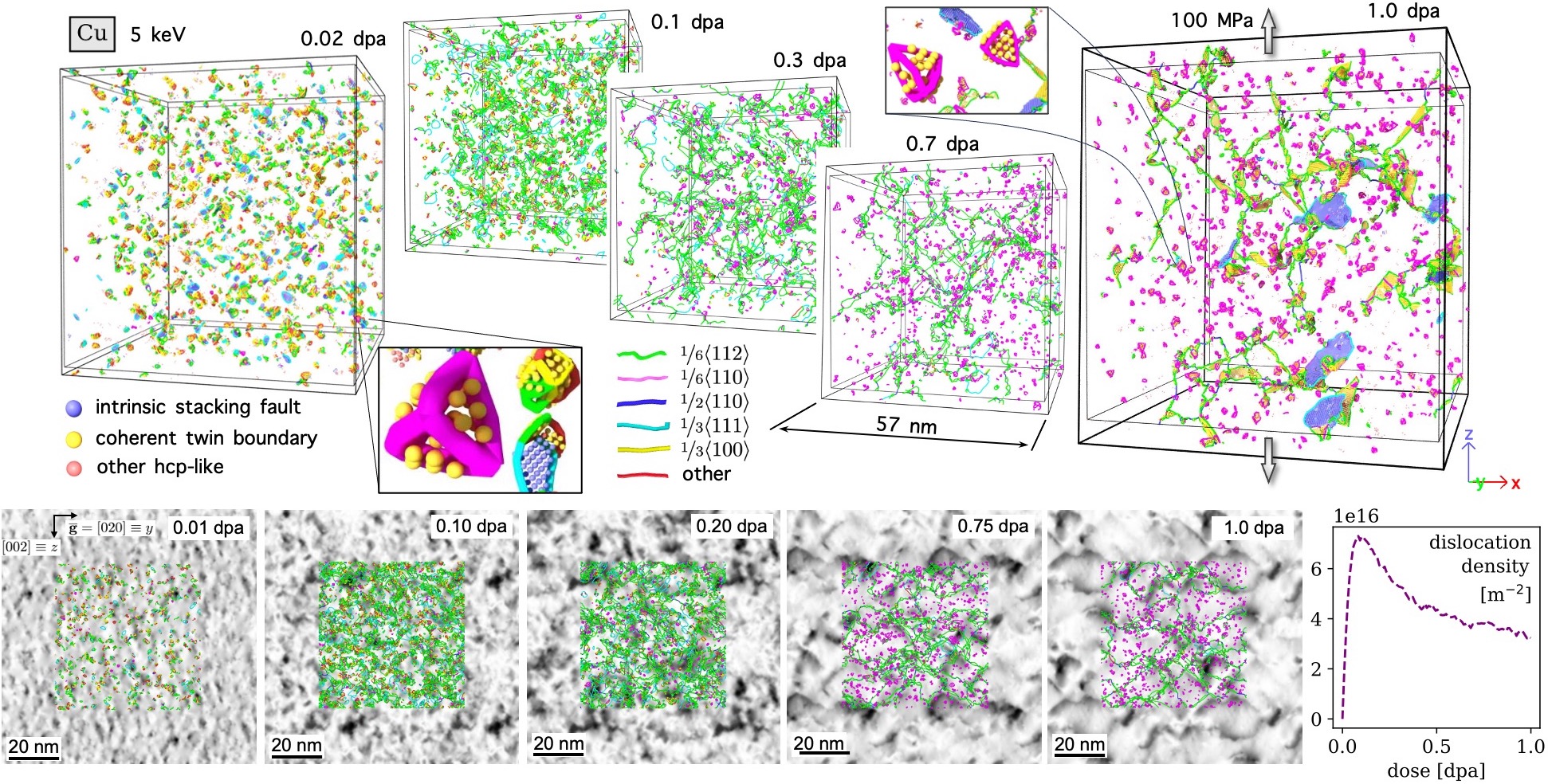}\label{fig:disloc_b} %
}\hfill
    \caption{Dislocation evolution under high-energy irradiation in sheared tungsten (a) and tensioned copper (b): isolated loops forming at very low dose evolve into a dislocation network that subsequently undergoes partial annihilation and simplification under continuing irradiation. This leaves larger dislocation loops and, in Cu, stacking fault tetrahedra (SFTs). Vacancies and self-interstitials are visualised, in W, in the low- and high-dose limits. About 90\% of the vacancies remain  homogeneously distributed while self-interstitials form small loops (inset) that coalesce and form the nearly complete crystallographic atomic planes. In Cu, fcc planar faults are visualised instead, showing a dispersed distribution of SFTs (insets) and a somewhat simpler network of mostly partial dislocations. Microstructural polarisation is apparent from a comparison of the instantaneous and initial simulation cells, shown by magnifying the deformations of the former by a factor of 20. In each dislocation analysis, the simulated dynamical bright-field TEM images are overlayed with dislocation lines from the atomistic simulation, where the simulation cells are periodically extended in the plane of the image ($\mathbf{g}=[002]$ for W and $\mathbf{g}=[020]$ for Cu). Defects smaller than approximately 2nm are not resolved by TEM \cite{Zhou2006,Mason2024}.}
    \label{fig:disloc} 
\end{figure*}

\begin{table*}[t]
\caption{Summary of effects of irradiation at {1\,dpa} in terms of the vacancy concentration $c_\textnormal{V}$, the volume change $\Delta V/V_0$, the von Mises strain increase $\Delta\varepsilon_{\textnormal{vM}}$ and the total dislocation density $\rho_\textnormal{disl}$. Here, $\Delta$ refers to the values computed taking the unirradiated, purely elastically loaded hydrostatic and deviatoric strained cells as references. The elastic strain is already offset in Fig.~\ref{fig:delta_V_a} while it is the difference between the first and the last points in Fig.~\ref{fig:delta_V_b} that is given here. The barely noticeable effect of the applied stress on the vacancy content and swelling is apparent, and the role of stress in generating the deviatoric distortions. Since different types of applied stress produce different von Mises stresses, values of $\Delta\varepsilon_{\textnormal{vM}}$  normalised by $\sigma_{\textnormal{vM}}$ are also shown for cases where $\sigma_{\textnormal{vM}}\ne 0$.}
\label{tab:summary}
\centering
\begin{tabularx}{\textwidth}{lXXXXc|XXXXc}\hline\hline
     & $c_\textnormal{V}$    & $\Delta V/V_0$  & $\Delta\varepsilon_{\textnormal{vM}}$  & $\Delta\varepsilon_{\textnormal{vM}}/\sigma_{\textnormal{vM}}$  & $\rho_{\textnormal{disl}}$ & $c_\textnormal{V}$    & $\Delta V/V_0$  & $\Delta\varepsilon_{\textnormal{vM}}$  & $\Delta\varepsilon_{\textnormal{vM}}/\sigma_{\textnormal{vM}}$ & $\rho_{\textnormal{disl}}$  \\
     & [\%] & [\%] & [\%] & [\%/GPa] & [10$^{16}$ m$^{-2}$] & [\%] & [\%] & [\%] & [\%/GPa] & [10$^{16}$ m$^{-2}$]\\ \hline
& \multicolumn{10}{c}{tungsten}\\
 & \multicolumn{5}{c}{90 eV} & \multicolumn{5}{c}{\SI{7}{\kilo\electronvolt}} \\
unstressed    & 1.49  &   0.89  &  0.38 & - & 0.43 & 0.51 & 0.31  &  0.38 & -  & 0.17\\
tension    & 1.45  &   0.87  &  1.22  &  3.46 & 0.23 & 0.51 & 0.31  &  0.55  &  1.55 & 0.40\\
compression    & 1.52  &   0.91  &  0.59  &  1.66 & 0.52 & 0.51 & 0.31  &  0.27  &  0.77 & 0.32\\
hydrostatic   & 1.54  &    0.91  &  0.52  & - & 0.59 & 0.51 & 0.31  &  0.39  & -  & 0.29  \\
shear    & 1.36   &    0.83  &  1.26  &  1.45  & 0.31 &  0.51 & 0.32  &  0.56  &  0.64 & 0.14\\
triple shear   & 1.48  &    0.88  &  1.71  &  1.14  & 0.41 & 0.53 & 0.32  &  0.76  &  0.51 & 0.24\\ \hline
& \multicolumn{10}{c}{copper}\\
 & \multicolumn{5}{c}{100 eV} & \multicolumn{4}{c}{\SI{5}{\kilo\electronvolt}} \\
unstressed    & 2.40  &   2.22  &  0.23 & - & 7.62 & 0.81 & 0.76  &  0.42 & -  & 3.46\\
tension    & 2.40  &   2.21  &  0.35  &  0.99 & 6.79 & 0.81 & 0.76  &  0.46  &  1.31 & 3.23\\
compression    & 2.39  &   2.20  &  0.37  &  1.05  & 7.00 & 0.79 & 0.74  &  0.44  &  1.24 & 3.23\\
hydrostatic   & 2.43  &    2.23  &  0.33  & -  & 7.34  & 0.82 & 0.76  &  0.19  & -    & 3.58 \\
shear    & 2.42   &    2.15  &  2.08  &  2.40  & 4.71 & 0.83 & 0.74  &  0.93  &  1.08 & 2.90\\
triple shear   & 2.42  &    2.16  &  1.18  &  0.79  & 5.09 & 0.79 & 0.73  &  0.59  &  0.39 & 3.06 \\ \hline\hline
\end{tabularx}
\end{table*}

\subsection{Defect evolution}
How do (i) the saturating and nearly stress-independent swelling and (ii) the strongly stress-dependent dimensional changes relate to the defect structures, representing the fundamental source of these changes? In Fig.~\ref{fig:disloc_a}, we show how dislocations, self-interstitial atom defects, and vacancies, identified here using the Wigner-Seitz method, evolve in W under shear stress. Vacancies form an increasingly denser homogeneous dispersion as their concentration rises. Self-interstitial atom defects rapidly start coalescing and forming interstitial-type prismatic dislocation loops with habit plane normal vectors pointing in the $[\overline 101]$ direction, which is orthogonal to the $[101]$ direction of the maximum principal applied stress. 

The observed difference between the evolution of vacancies and self-interstitial atom defects can be partially attributed to the lower mobility of vacancies that have a much higher thermal migration energy barrier of \SI{1.7}{\electronvolt} than the barrier for migration of self-interstitial defects that is as low as \SI{0.01}{\electronvolt} \cite{Swinburne2017}. Similar behaviour is observed in Cu, where the difference between the migration energy barriers is less dramatic, with the vacancy migration energy being close to 0.7 eV and self-interstitial atom migration energy being close to \SI{0.09}{\electronvolt} \cite{Andersson2004, Zhang2022}. This might also explain why the microstructural evolution in Cu occurs slower than in W, extending into the dose interval beyond {1\,dpa}. Cu, as illustrated in Fig.~\ref{fig:disloc_b}, showing the evolution of Cu under tension, exhibits the formation of dispersed stacking fault tetrahedra (SFTs), in agreement with experimental observations \cite{Schaublin2005}. Copper is also characterised by the overall slower evolution of the dislocation network as a function of dose.

A widely used tool for observing the microscopic effects of radiation damage is transmission electron microscopy (TEM). It is therefore instructive to visualise radiation defects as they would appear in TEM images. Fig.~\ref{fig:disloc} illustrates images of defects responsible for the observed mechanical response of the materials. These images were produced directly from the coordinates of atoms in the simulation cell using a two-beam dynamical diffraction TEM image simulation code \cite{Mason2024}. Fig.~\ref{fig:disloc} also shows dislocations, identified using the dislocation extraction algorithm (DXA) implemented in Ovito \cite{Stukowski2009}. In agreement with experimental TEM observations \cite{Wielunska2022, Wang2023}, in the very low dose limit ($\lesssim$0.0{1\,dpa}), only the isolated dislocation loops are observed, which then rapidly grow to form a tangled network of dislocations at the dose of order $\sim$0.1 dpa. Further irradiation gives rise to a substantial simplification of the network, in agreement with high dose cascade simulations \cite{Boleininger2022}. Qualitatively, the dislocation evolution observed by TEM in bcc W \cite{Wang2023} is also similar to observations of defect accumulation in ion-irradiated bcc-Fe \cite{Hernandez2008}. Note that nanometre-size loops and SFTs are not always visible in the two-beam TEM images \cite{Zhou2006}. 

The dislocation density and vacancy concentration increase rapidly over the initial dose interval of {$\sim$0.1\,dpa} under high-energy irradiation, and over the {$\sim$0.2\,dpa} dose interval under low-energy irradiation. Once a threshold dose is exceeded, the dislocation objects that initially formed as isolated prismatic dislocation loops, merge into an extended interconnected dislocation network. Further irradiation gives rise to a slow partial simplification of the network, which occurs faster in W than in Cu. The pattern of evolution, found here using the simulations based on the molten spheres algorithm, is consistent with the earlier experimental and modelling work \cite{Wang2023, Derlet2020, Boleininger2022}. The vacancy concentration, computed using an algorithm formulated in \cite{Mason2021}, approaches saturation at a dose of order {1\,dpa}. Notably, the above conclusions are almost independent of the applied stress, see also Supplementary Material, in agreement with experimental observations of unstressed and tensioned W \cite{Ponsoye1971}. The patterns of variation of swelling and vacancy concentration as functions of dose stay the same irrespective of the applied stress, and both swelling and vacancy concentration similarly approach a plateau, whereas a somewhat higher dose scale, characterising the evolution of dislocation density, is better correlated with the dose scale characterising the variation of the deviatoric part of the total strain. 

The saturated microstructural state of the material, forming in the limit of high exposure close to 1 dpa, provides a good representation of microstructure forming under low-temperature irradiation, and it is different from the microstructure forming at higher temperatures, where it is often dominated by the thermal coalescence of vacancies into voids and by the segregation of transmutation products, not considered here.  Tab.~\ref{tab:summary} summarises results of {1\,dpa} simulations, carried out for all the types of applied stress, highlighting the marked role of the effective recoil energy, with about three times as much swelling and as many vacancies forming for the conditions resembling electron irradiation than for conditions representative of high-energy heavy-ion or neutron irradiation. These values agree with cascade simulations of W \cite{Boleininger2023}, with a slight overestimation of order {20 to 25\,\%}, confirming that the defect content can vary significantly if the same dose is delivered by particles with different effective recoil energies. The low-energy irradiation also gives rise to a stronger polarised eigenstrain $\omega_{ij}$, albeit the difference between the polarisations computed for the low and high PKA energies is not as dramatic as for the volumetric swelling. Overall, we see that although the defect \emph{content} is barely affected by stress, the reorganisation of self-interstitials via dislocation motion results in a large variation of the material's irradiation creep response even in the athermal regime.

\subsection{Strain and eigenstrain from irradiation}
MD simulation boxes are defined by three supercell vectors. Let us define a matrix $\mathbf{A}$ whose columns are the supercell vectors. A change in length of the vectors results in tensile or compressive strains, a change in angle between them results in shear strains, or tilting of the box. So far we calculated total strains from the deformation of the MD box, or in other words from changes of matrix $\mathbf{A}$ using the definitions in the Supplementary Material. The total strain is observable and relevant for creep \cite{Feichtmayer2024} and for macroscopic effects caused by gradients of swelling \cite{Reali2022}. In the literature, a \emph{lattice} strain is also defined \cite{Lorentzen2003, Karato2009}, to measure strain from changes in interplanar distances. In practice, this second observable strain is measured from a diffraction pattern. Lattice strain in irradiated materials \cite{Mason2020} is important because defects respond to local lattice expansion or contraction, which affect, among a variety of other phenomena, the stress driven diffusion \cite{Ma2019} and deuterium retention. To define lattice strain, we must define the plane spacings. Let us define a second matrix, $\mathbf{B}$, whose columns are the primitive lattice vectors.
Under periodic boundary conditions, continuity requires that a supercell vector is an integer number of repeats of primitive unit cells, i.e. that
\begin{equation}\label{eq:sup_latt}
    \mathbf{A} = \mathbf{B} \mathbf{N}
\end{equation}
where $\mathbf{N}\in\mathbb{Z}^{3\times3}$. 
For the W example, this equation links the cubic supercell to 200$\times$200$\times$200 conventional bcc unit cells with lattice parameter $a$, as 
\begin{equation*}
    \small
    \left( \begin{array}{ccc}
        200a    &   0       &   0   \\
        0       &   200a    &   0   \\
        0       &   0       &   200a    
        \end{array} \right)
    =
    \left( \begin{array}{ccc}
        ^{-a}\!/\!_2    &   ^a\!/\!_2     &   ^a\!/\!_2 \\
        ^a\!/\!_2     &   ^{-a}\!/\!_2    &   ^a\!/\!_2 \\
        ^a\!/\!_2     &   ^a\!/\!_2     &   ^{-a}\!/\!_2    
        \end{array} \right)
    \left( \begin{array}{ccc}
        0       &   200     &   200   \\
        200     &   0       &   200   \\
        200     &   200     &   0    
        \end{array} \right).
\end{equation*}
Note that the columns of the periodicity matrix $\mathbf{N}$ are the number of primitive cell vector steps required to cross the supercell and return to the starting point across one of the three supercell faces, and so are analogous to Burger's circuits modulo the periodicity. The number of lattice sites is $\mathrm{det}(\mathbf{N})$ times the number of motif points per primitive cell, equal to 1 for bcc and fcc.

Initially, isolated defects cause a small amount of strain.
$\mathbf{N}$ is unchanged, and the supercell total strain and the lattice strain are constrained to remain the same by \eqref{eq:sup_latt}. 
As dislocation loops coalesce and grow, with the simulation cell size staying constant, the number of primitive cell vector steps required to cross the supercell can change, and so $\mathbf{N}$ changes.
As an example, an interstitial loop comparable in size to the supercell will start to approximate a complete atomic plane \cite{Mason2020, Boleininger2022}.
This means that some interstitial atom defects should be re-categorised as perfect crystal atoms and vacancies must now outnumber the self-interstitials to preserve the atom count while increasing the lattice site count \textemdash this implies substantial swelling.
The consequent change in $\mathbf{N}$ \emph{increases} $\mathrm{det}(\mathbf{N})$.
Here we have also observed changes in $\mathbf{N}$ which \emph{preserve} $\mathrm{det}(\mathbf{N})$ \textemdash these are pure slip events.

\begin{figure}[t]
  \includegraphics[width=0.9\columnwidth]{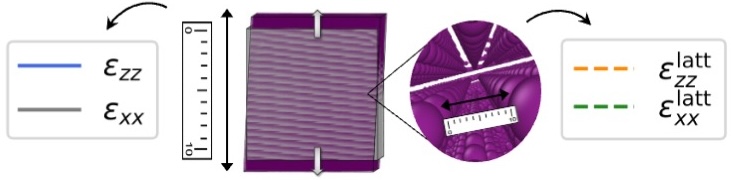} %
\subfloat[]{%
\includegraphics[width=0.95\columnwidth]{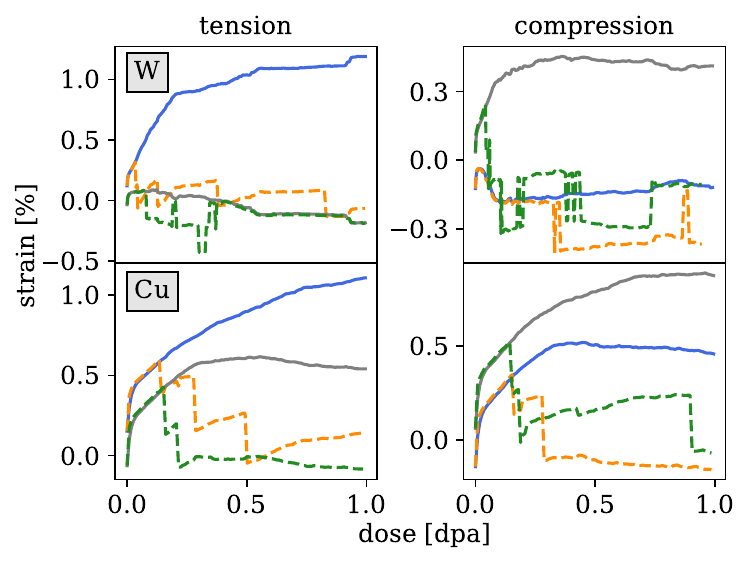}\label{fig:lattice_strain_a}  %
}\hfill
\subfloat[]{%
  \includegraphics[width=.85\columnwidth]{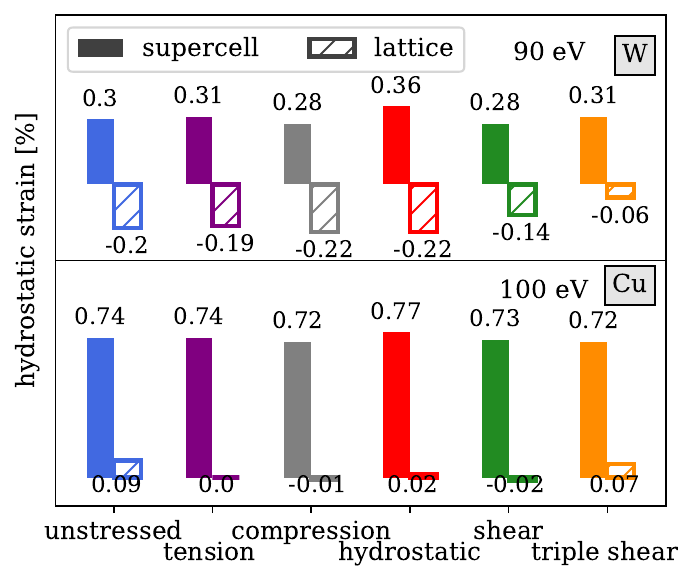}\label{fig:latt_hist}\label{fig:lattice_strain_b} %
}\hfill
    \caption{Swelling and creep of MD box\textemdash the concepts used throughout the paper\textemdash are computed as if measuring the distortion of the entire ensemble of atoms with a ruler. Another, independent, way involves measuring changes in the interatomic spacings with, in the same metaphor, an atomic ruler, obtaining a lattice strain. We can see that initially the two definitions produce the same values but rapidly, if the material undergoes excessive deformation, the lattice strain collapses and remains low in absolute value to minimise the increase in the strain energy. For clarity, here we show two of the strain components.}
    \label{fig:lattice_strain} 
\end{figure}
Changes in $\mathbf{N}$ imply that lattice and supercell strain become decoupled as is clearly visible in Fig.~\ref{fig:lattice_strain_a}, where strain components during the tensile and compressive simulations are shown. 
The steps seen in the lattice strain in Fig.~\ref{fig:lattice_strain_a} are of order $^a\!/\!_{2 L}$, where $L$ is the supercell side length, and are a finite size effect due to the integer nature of $\mathbf{N}$ for a periodically translated simulation cell. The step height could be reduced by averaging over independent simulations or by increasing the system size $L$.
In a real crystal grain, the constraint on periodicity is lifted, but still there will be two strain tensors because the crystal lattice has internal degrees of freedom \textemdash a full description of the size of the grain requires both the lattice strain and the lattice sites, and the latter is changed by swelling and by slip \cite{Ma2024}. Fig.~\ref{fig:lattice_strain_b} shows moderate \emph{negative} hydrostatic lattice strain at high dose in all tungsten simulations, although the supercell has a larger \emph{positive} expansion. Lattice strains in copper are barely noticeable. The full strain tensor whose hydrostatic component is shown in Fig.~\ref{fig:lattice_strain_b}, and additional details on the calculation of the lattice strain are given in the Supplementary Material. 

We now return to considering the total supercell deformations that can inform higher-scale models of swelling and creep. Experimental data suggest that in the dose interval up to {0.1-0.5\,dpa}, the low-temperature creep strain rate is high and is proportional to the applied stress and the dose rate $\Dot{\phi}$ \cite{Hoffelner2010, Scholz2000},
\begin{equation}
    \Dot{\varepsilon}=C\cdot\sigma\cdot\Dot{\phi}.\label{creep_law}
\end{equation}
This is consistent with the exponential stress relaxation law \cite{Causey1980, Grossbeck1991} ${\sigma=\sigma_0\textnormal{exp}(-C\cdot\phi/E)}$ that can be rationalised by relating the increments of strain and stress through  $\textnormal{d}\varepsilon=E\textnormal{d}\sigma$, where $E$ is the Young modulus. The creep compliance coefficient $C$ is of order $10^{-5}$~MPa$^{-1}$dpa$^{-1}$ in nickel alloys (\SI{300}{\celsius} \cite{Scholz2000}) and steel \cite{Grossbeck1991}. At high dose, the low temperature irradiation creep eventually saturates \cite{Scholz2000, Ponsoye1971,Luzginova2011} and hence the trend given by Eq. (\ref{creep_law}) cannot be maintained.

\begin{figure}[t]
  \includegraphics[width=0.9\columnwidth]{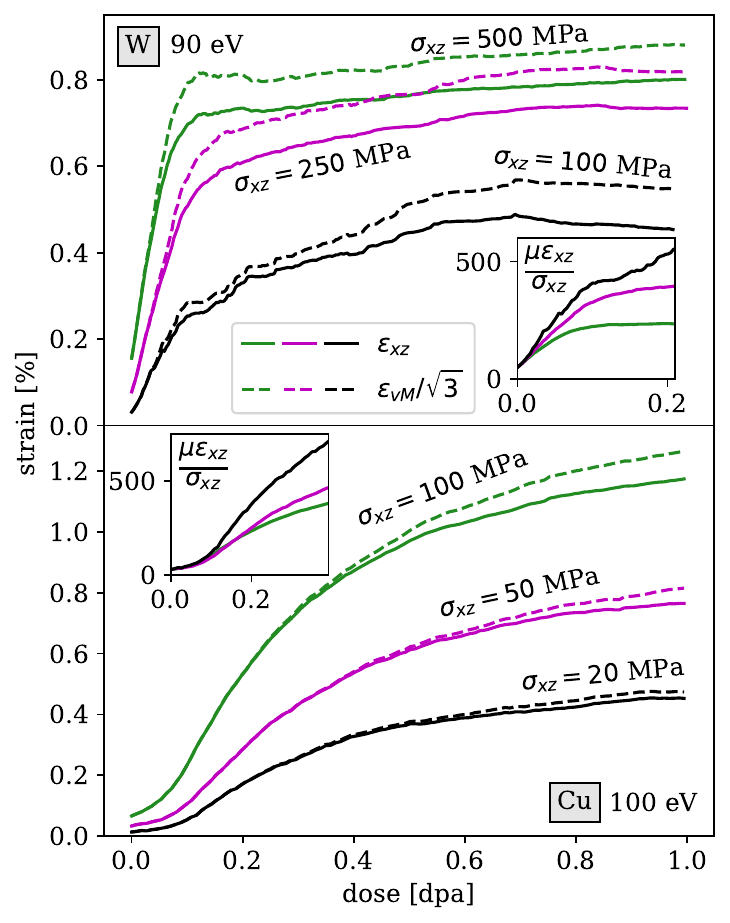} %
    \caption{The magnitude of applied shear stress affects the polarisation of swelling: an increase of $\sigma_{xz}$ leads to a larger value of $\varepsilon_{xz}$, whereas the volumetric swelling is affected to a far lesser degree. The relative effect is higher for smaller stresses, as shown in the insets, e.g. reducing $\sigma_{xz}$ by {50\,\%} causes a drop in  $\varepsilon_{xz}$ at {1\,dpa} of {8\,\%} in W and {35\,\%} in Cu (cf. the shear modulus $\mu$ for W is  \SI{161}{\giga\pascal} whereas it is \SI{75}{\giga\pascal}  for Cu). Finally, the von Mises strain $\varepsilon_{\textnormal{vM}}$ stems largely from the contribution of the $xz$ strain component that is the same as the applied stress.}
    \label{fig:3stress} 
\end{figure}

We have explored the sensitivity of the induced total strain to the applied stress under shear. Fig.~\ref{fig:3stress} shows how $\varepsilon_\mathrm{xz}$ in the cell, the dominant component of strain contributing to $\varepsilon_{\textnormal{vM}}$ in this case, is affected by the applied shear stress. Increasing $\sigma_\mathrm{xz}$ from 100 to 250 and then to 500 MPa in W under low-energy irradiation results in the increase of $\varepsilon_{\textnormal{vM}}$ at {1\,dpa} from 0.9\,\% to 1.4\,\% and to {1.5\,\%}. In Cu, the increase of $\sigma_\mathrm{xz}$ from 20 to 50 to 100 MPa produces the variation of $\varepsilon_{\textnormal{vM}}$ at {1\,dpa} of 0.8\,\%, 1.4\,\% and {2.2\,\%}. Swelling is nearly unaffected, remaining close to 0.9\,\% and {2.1\,\%} in W and Cu, respectively, for the three shear stress magnitudes. This supports the hypothesis that stress does not  affect the defect content but rather how the defects evolve, suggesting that the driving force for microstructural evolution in the athermal limit stems primarily from the internal and applied stresses. For a given principal stress direction, there is an `ideal polarised microstructural state'. The greater normalised effect of a smaller stress, see the insets in Fig.~\ref{fig:3stress}, suggests that increasing the stress enables a faster and more complete re-configuration of the defects, but even a small stress can slowly drive the irradiation creep. Finally, note that even shear stress of order of {0.05\,\%} of the respective shear modulus clearly affects the deviatoric part of the total strain. Before approaching saturation, below about 0.{1\,dpa}, we found $\varepsilon\propto\sigma$, as shown in the insets in Fig.~\ref{fig:3stress}. The creep compliance is of order of $10^{-5}$ to $10^{-4}$ for W and $10^{-5}$~MPa$^{-1}$dpa$^{-1}$ for Cu, remaining lower for the higher-energy irradiation. This suggests that athermal stress-driven defect reconfiguration is the main factor responsible for the radiation creep at low temperature. This conclusion was also reached in Ref.~\cite{Feichtmayer2024}.

\subsection{Strain and deformation decomposition}
We started the study by noting that the irradiation-induced eigenstrain tensor $\omega_{ij}$ is sensitive to the stress applied to the material under irradiation. 

\begin{figure}[t]
  \includegraphics[width=.5\textwidth]{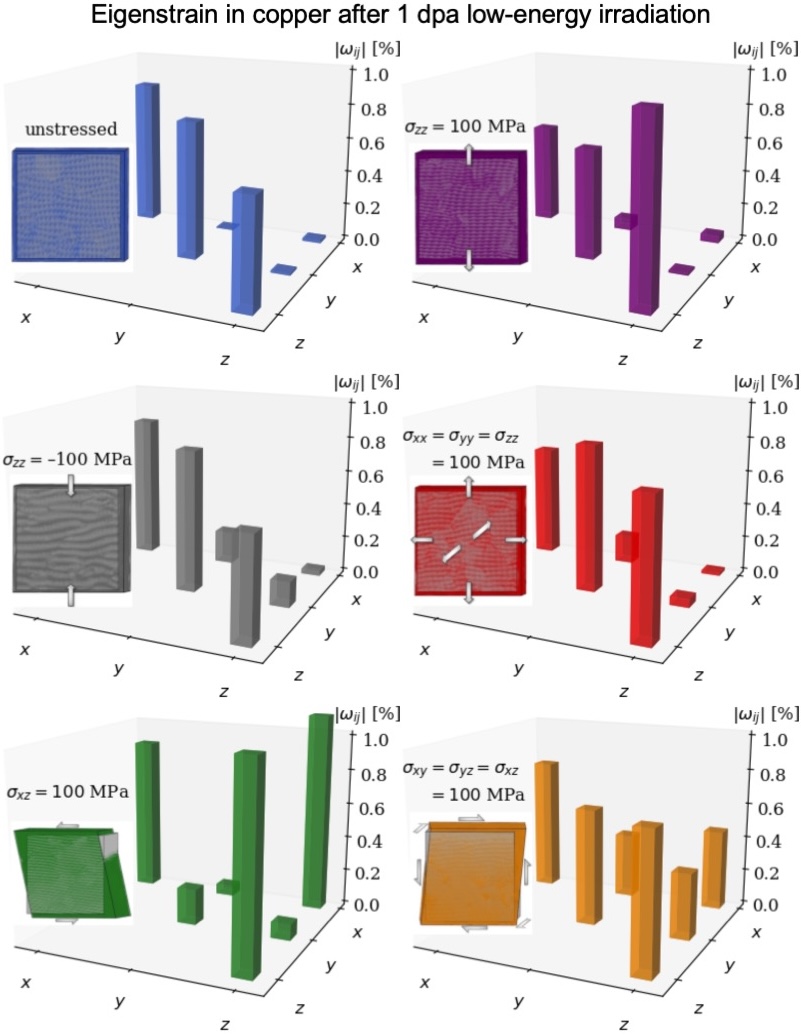} %
    \caption{During the radiation exposure to {1\,dpa} under six different constant external stresses, copper responds by creeping anisotropically. Relaxing the external stress reveals the underlying tensorial eigenstrain. Although here we show the absolute values of its components, all the diagonal components are positive. Qualitatively, the behavior of tungsten is similar.}
    \label{fig:eigenstrain} 
\end{figure}

To show that this is indeed the case, we recall the definition of the total strain $\varepsilon_{ij}=\omega_{ij}+e_{ij}$ \cite{Mura2013, Reali2022}, where $e_{ij}=C_{ijkl}\sigma_{kl}$. Here, the eigenstrain term $\omega _{ij}$ arises solely from radiation defects. To separate the eigenstrain $\omega_{ij}$ from elastic strain, we have unloaded all the simulated 1\;dpa microstructures by setting $\sigma_{kl}=0$. Now, the total strain in the simulation cell is identical to eigenstrain, averaged over the cell. During the unloading, only a small elastic recovery of defect structures is observed and defects do not appreciably move. In this way, we separate the elastic and irreversible strains. The irreversible strain is accumulated solely by defect generation and recrystallisation, and there is no contribution from thermally activated dislocation-mediated plasticity. $\omega_{ij}$ is markedly anisotropic and remains polarised by the stress that was acting during the exposure even after this stress has been relaxed. Visually, this is presented for the case of low-energy irradiation of Cu in Fig.~\ref{fig:eigenstrain}.

To describe a general case involving a combination of elastic, plastic, and radiation-induced deformations, we propose to use a three-fold decomposition of the deformation tensor 
\begin{equation} \label{eq:CP_3decomposition}
    \mathbf{F}=\mathbf{F}^e\mathbf{F}^p\mathbf{F}^\omega.
\end{equation}
In the CP formulations available in literature, similar decompositions are used for treating the effects of thermal expansion \cite{Musinski2015}. Our treatment therefore becomes fundamentally consistent with the dislocation glide-based formalism for irradiation creep, even though we do not discuss effects of thermally-activated dislocation glide here. Similarly, in the presence of thermally-assisted defect recombination there are fewer defects available for driving creep and swelling than those analysed in this study. The fundamental features of the models remain the same, each applicable in its own temperature ranges. The U-shaped temperature trends exhibited by swelling and irradiation creep, saturating at low-temperature and exhibiting breakaway growth at high-temperature, are reflected in the $\mathbf{F^\omega}$ having the characteristics explored in this work and in the thermally-activated nature of $\mathbf{F}^p$ \cite{Yu2022}.  

\begin{figure*}[t]
\subfloat[]{%
  \includegraphics[width=.67\textwidth]{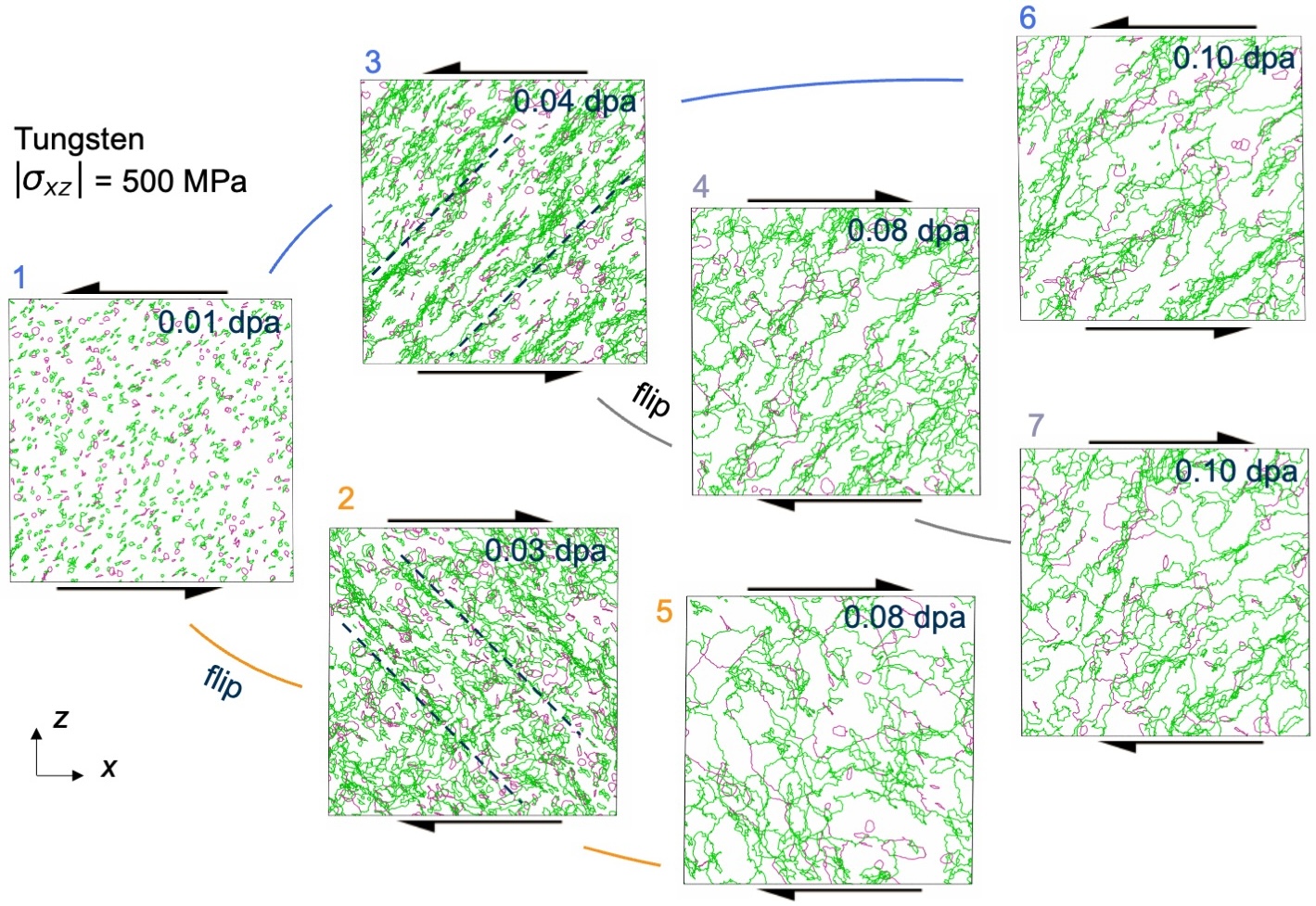}\label{fig:flip_DXA} %
}\hfill
\subfloat[]{%
  \includegraphics[width=.31\textwidth]{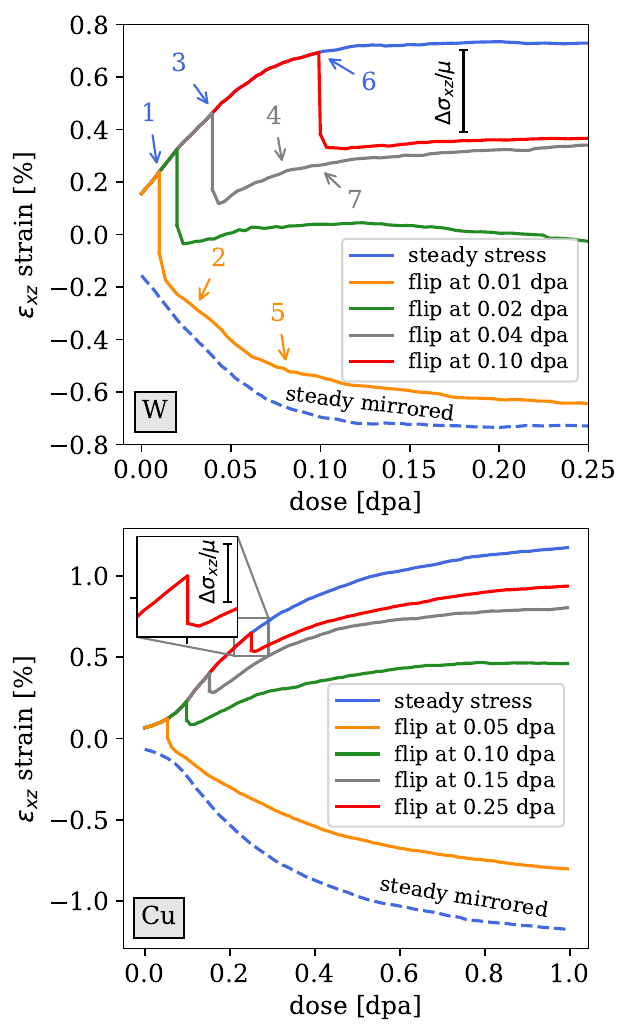}\label{fig:flip_strain} %
}\hfill
    \caption{Irradiation creep of W and Cu begins under pure shear stress before reversing the sign of the stress at various levels of exposure. The shear strain undergoes instantaneous elastic recovery but creep in the direction opposite to that of the initial shear occurs only after very modest exposure. The W dislocation analysis (a) shows how under steady stress, the dislocation network is clearly aligned one way (dashed lines), and aligned the opposite way after the stress reversal at 0.01\;dpa. After flipping the direction of stress at 0.04\;dpa, however, the network still retains memory of its alignment. The response is markedly irreversible as highlighted by the strain evolution shown in (b).}
    \label{fig:flip} 
\end{figure*}

For engineering applications, it is important to assess the reversibility of the irradiation-induced creep deformation. In a tensioned bolt the loading is uniaxial, but reactor components in general operate under thermo-mechanical tri-axial cycling loads. How does the material respond if further loading is applied to an already irradiated material previously subjected to creep conditions? Experiments on steels \cite{Garner2009} and aluminium \cite{DaFonseca2023} suggest that\textemdash at least initially\textemdash the pre-existing irradiation-induced dislocation microstructure is retained in the materials, and this influences its creep response. To investigate this in detail, we extended the low recoil energy simulations under simple shear conditions, where we flipped the sign of shear stress $\sigma_{xz}$ at different exposures and tracked the deformation $\varepsilon_{xz}$ of the simulation cell, as illustrated in Fig.~\ref{fig:flip}. By doing so, we reverse the global, average elastic stress but this is not followed by a noticeable re-arrangement of the defects because of the lack of thermal activation. We observe a jump in the deformation very close to the elastic threshold $\Delta\sigma_{xz}/\mu=2\sigma_{xz}/\mu$, followed by a variety of irradiation creep response modes, depending on the dose at which the applied stress changed sign. At a very low dose, the strain increments also responded by changing the sign, even though the material never reached the $\varepsilon_{xz}$ expected if the simulation had begun with the flipped shear stress from the start. But the response become more complex if the applied stress is reversed at a dose when the dislocation network had already formed and the finely dispersed distribution of vacancies had already developed, bringing the material close to a dynamic high dose state, where the high density of vacancies ensures that the newly formed self-interstitial defects recombine almost immediately after their formation. An examination of the dislocation structure in W shows that the polarised dislocations flip the initial orientation only if the stress change occurred at or before 0.01\;dpa. This corroborates the hypothesis proposed in Ref. \cite{DaFonseca2023} about the dominance of the stress-induced nucleation mechanism rather than absorption. 
The reason is expected to be that \emph{locally} the stress produced by the already established defect and dislocation configuration dominates over the applied stress. Since this defect and dislocation configuration does not appreciably respond to the reversal of the applied stress, the subsequent evolution is predominantly driven by the microstructure that formed before the stress is reversed. This also suggests that \textemdash at least at low temperature \textemdash the irradiation creep is largely irreversible. We note that the model developed here describes the athermal limit where the pre-existing defects do not move through thermal activation. This stabilises the microstructure and maximises the irreversibility of the creep response. 
\section{Summary and conclusions}
High-dose irradiation of tungsten and copper irradiated under a variety of externally applied stress states was simulated atomistically using a novel molten spheres algorithm. The algorithm efficiently replicates the fact that collision cascades cause a localised melting, with the molten volume increasing as a function of the recoil energy of the atom initiating the cascade. This enables simulating low and high energy irradiation through the insertion of locally molten regions that subsequently recrystallise through the use of energy minimisation. The process simultaneously creates damage and erases it by recrystallising some of the pre-existing defects, mimicking a known genuine feature of irradiation \cite{Nordlund1997}.  Under the applied external stress, irradiation causes the formation and evolution of defects as well as changes in the atomic supercell volume \emph{and} shape. The vacancy content is almost insensitive to the applied stress and saturates at about 0.5\% in W and 0.8\% in Cu under high-energy irradiation, and at about 1.4\% in W and 2.2\% in Cu under low-energy irradiation. The change of simulation cell volume, i.e. swelling, is almost unaffected by stress. The deviatoric part of the total strain, related to radiation creep, is highly sensitive to the applied external stress. Under tension or shear stress, the irradiation creep strain polarises and becomes aligned with the applied stress tensor. The magnitude of hydrostatic and deviatoric strains in the high radiation exposure limit are much higher than the initial elastic deformations, reaching 0.5-2\;\% at {1\;dpa}. The hydrostatic component of the total strain is a measure of athermal swelling whereas the deviatoric part of the total strain is a measure of athermal irradiation creep. Both are present in applications and should be included in the finite element models, through the decomposition given by Eq.~\eqref{eq:CP_3decomposition}. In the low temperature applicability regime, we show how the eigenstrain polarises irreversibly under applied stress of different tensorial nature, magnitude, of steady or reversing character, with immediate relevance for reactor design and operations.
\bibliography{main}
\noindent\textbf{Acknowledgements}\\
{\footnotesize
The authors acknowledge discussions with A.E. Sand that stimulated this study and with A.R. Warwick. This work has been carried out within the framework of the EUROfusion Consortium, funded by the European Union via the Euratom Research and Training Programme (Grant Agreement No 101052200 — EUROfusion) and
was partially supported by the Broader Approach Phase II agreement under the PA of IFERC2-T2PA02. This work was also funded by the EPSRC Energy Programme (grant number EP/W006839/1). To obtain further information on the data and models underlying the paper please contact PublicationsManager@ukaea.uk. Views and opinions expressed are however those of the authors only and do not necessarily reflect those of the European Union or the European Commission. Neither the European Union nor the European Commission can be held responsible for them. This study was performed using resources provided by the Cambridge Service for Data Driven Discovery (CSD3).
}
\end{document}